\documentclass[prb,twocolumn,letterpaper,amsmath,amssymb,nobalancelastpage,preprintnumbers]{revtex4}

\usepackage{graphicx}% Include figure files
\usepackage{dcolumn}% Align table columns on decimal point
\usepackage{bm}% bold math
\usepackage{float}
\usepackage{textcomp}
\usepackage{amssymb}
\usepackage{amsmath}
\begin{document}

\preprint{Accepted for publication in Applied Physics Letters}
%\linenumbers
\title{Low-voltage organic transistors and inverters\\ with ultra-thin fluoropolymer gate dielectric}

\author{M. P. Walser}
  \email{walserma@phys.ethz.ch}
\author{W. L. Kalb}
\author{T. Mathis}
\author{B. Batlogg}%

\affiliation{\footnotesize Laboratory for Solid State Physics, ETH Zurich, 8093 Zurich, Switzerland}

%\date{\small \today}% It is always \today, today,

%\pacs{73.20.-r, 73.40.-c, 73.61.Ng, 73.61.Ph, 77.84.Jd}% PACS, the Physics and Astronomy
%73.20.-r Electron states at surfaces and interfaces
%73.40.-c Electronic transport in interface structures
%73.61.-r Electrical properties of specific thin films
%73.61.Ng Insulators
%73.61.Ph Polymers; organic compounds
%77.84.-s Dielectric, piezoelectric, ferroelectric, and antiferroelectric materials

%\keywords{low-voltage organic field-effect transistor, inverter, organic thin-film transistor, ultra-thin fluoropolymer gate dielectric, Cytop, electrical stability, gate bias stress}
%Use showkeys class option if keyword

\begin{abstract}
\begin{minipage}{0.8\textwidth}
We report on the simple fabrication of hysteresis-free and electrically stable organic field-effect transistors (OFETs) and inverters operating at voltages $<$1-2\,V, enabled by the almost trap-free interface between the organic semiconductor and an ultra-thin ($<$20\,nm) and highly insulating single-layer fluoropolymer gate dielectric (Cytop). OFETs  with PTCDI-C$_{\mathrm{13}}$ (N,N'-ditridecyl\-perylene-3,4,9,10-tetra\-carboxylic\-diimide) as semiconductor exhibit outstanding transistor characteristics: very low threshold voltage (0.2\,V), onset at 0\,V, steep subthreshold swing (0.1-0.2\,V/decade), no hysteresis and excellent stability against gate bias stress. It is gratifying to notice that such small OFET operating voltages can be achieved with the relatively simple processing techniques employed in this study.
\end{minipage}
\end{abstract}

\maketitle

% include figure 1 - use [scale=1.1] for twocolum print

% figure 1
\begin{figure}
\includegraphics[scale=1.1]{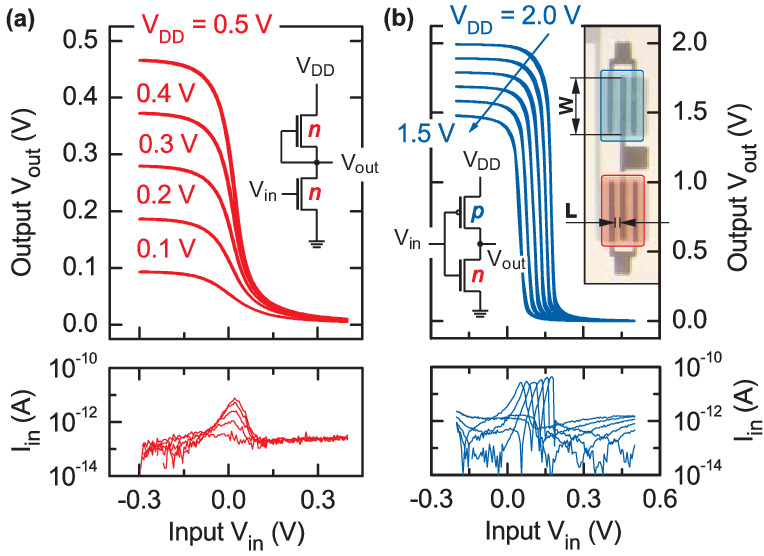}
\caption{\footnotesize \label{fig:ntypeInv} (Color online) Organic inverters with an 19\,nm thin Cytop dielectric: (a)~$N$-type PTCDI-C$_{\mathrm{13}}$ inverter with depletion load and (b)~a complementary PTCDI-C$_{\mathrm{13}}$/pentacene inverter.  The maximal gain (derivative of the static inverter characteristic) was -4.5 for $n$-type PTCDI-C$_{\mathrm{13}}$ inverters and -80 for complementary PTCDI-C$_{\mathrm{13}}$/pentacene inverters. Insets: Circuit layout of $p$- and $n$-type transistor combination. Optical micrograph (colored): Device geometry. Lower panels: The gate leakage current is less or similar to 1\,pA (forward scan shown). 
}
\end{figure}

Organic electronics is expected to enable novel applications, such as large area electronics on flexible and transparent substrates or printed intelligent identification tags.
The electrical current in OFETs is mainly confined to the molecular layers in close proximity to an adjacent insulating layer, the gate dielectric. A morphologically and chemically ideal interface between semiconductor and dielectric is free of charge carrier traps. However, charge transport and device stability are adversely affected by imperfections and contaminants, in particular by electrically active chemical groups, often related to water\cite{Chua2005,Goldmann2006,Pernstich2006,Mathijssen2008}.

Of particular interest for organic electronics is the simple fabrication of organic thin-film transistors (OTFTs) with low operating voltage and high operational stability.
High performance and low operating voltage have been achieved by the fabrication of self-assembled molecular monolayer\cite{Fontaine1993,Halik2004,Klauk2007} and multilayer dielectrics\cite{Yoon2005a,DiBenedetto2009}. Among many novel polymeric dielectrics, such as cross-linkable polymers\cite{Chua2004,Facchetti2005,Jang2006,Kim2008a,Noh2009,Roberts2009} and polymer blends\cite{Chen2004,Schroeder2005,Yoon2005}, fluoropolymers are particularly promising to meet various requirements: (i)~excellent insulation (low gate leakage currents at high electric fields), (ii) high capacitance per unit area (simple and reproducible fabrication of very thin and flat films), (iii)~high chemical stability and high water repellency (low interface trap-density, high operational stability and high mobility in combination with both $p$- and $n$-type organic semiconductors). The commercially available fluoropolymer Cytop\texttrademark\, (Asahi Glass Japan) is an amorphous, optically transparent polymer with a low permittivity (2.1) and has initially been used as gate dielectric in OFETs by Veres et al.\cite{Veres2004}. It is highly water repellent and yields OFETs of very high electrical quality and very high resistance against gate bias stress\cite{Kalb2007a}. Cytop is an excellent electrical insulator\cite{Kalb2007a}, which was also shown for very thin Cytop films, used as gate insulator for pentacene TFTs\cite{Umeda2008}. Both $p$- and also $n$-type OFETs have been demonstrated on a standard SiO$_{2}$ dielectric coated with a Cytop layer\cite{Walser2009}. The absence of hydroxyl groups and a very high water repellence are preferable interface properties, particularly due to the possibility of electron trapping at the interface by hydroxyl groups\cite{Chua2005}. Inorganic dielectrics with a polymer coating or a SAM surface treatment are widely used to fabricate low-voltage complementary inverters\cite{Vusser2006,Kitamura2007,Zhang2009}. 

In this study, we exploited both the highly desirable chemical properties and the excellent insulating properties of ultra-thin Cytop films used as single-layer gate dielectric. In contrast to prevailing assumptions on polymer insulators in general, here we show that sub-20\,nm thin fluoropolymer gate dielectrics can provide for high breakdown fields and low current leakage, even if spin-coated over patterned and rather rough bottom gate electrodes. We demonstrate organic thin-film transistors (OTFTs) with excellent electrical characteristics and with low operating voltage, enabling inverter operation below 1-2\,V (Fig.\,\ref{fig:ntypeInv}). A few tenths of a volt (0.3-0.6\,V) at the input ($V_{\mathrm{in}}$) of an organic inverter are sufficient for switching the output ($V_{\mathrm{out}}$) between logic high ($V_{\mathrm{DD}}$) and low signals (0\,V).  

\begin{figure}
\includegraphics[scale=1]{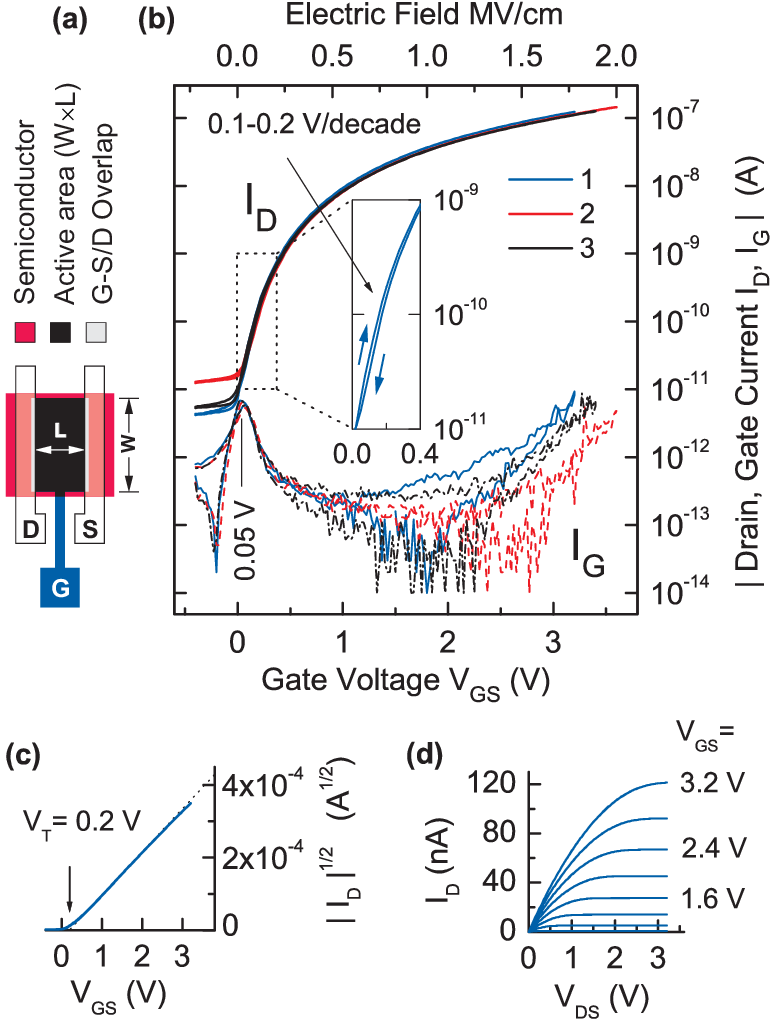}
\caption{\footnotesize \label{fig:trans18nmStat} (Color online) Performance of PTCDI-C$_{\mathrm{13}}$ TFTs with an 18\,nm thin Cytop dielectric: (a)~Schematic of the TFT geometry used in this study. (b)~Transfer characteristics for three typical devices (saturation regime). The gate current is below 10\,pA at 2\,MV\!/cm. Inset: The onset is essentially at 0\,V, the hysteresis is extremely small ($\approx$\,0.02\,V) and the subthreshold swing very steep (0.1-0.2\,V/decade). (c)~The threshold voltage is as low as 0.2\,V. (d)~Output characteristics are nearly ideal. 
}
\end{figure}

Prior to spin-coating the Cytop layer, we deposited $\approx$\,15\,nm of Al at a rate of 0.5-2\,\AA/s as bottom gate electrode onto glass substrates. All metals and organic semiconductors were thermally evaporated through shadow masks at a base pressure near 3$\times$10$\mathrm{^{-6}}$\,mbar. For 18-20\,nm thin Cytop films, Cytop CTL-809M was dissolved in CT-Solv.180 in the ratio $\sim$\,1:10, spin-coated onto the patterned gate electrodes at 500\,rpm for 10\,s, then at 2000\,rpm for 20\,s, and afterwards dried on a hot plate for 30\,min at each 55, 90 and 120\,\textdegree{}C. Approximately 50\,nm PTCDI-C$_{\mathrm{13}}$ and pentacene (Sigma Aldrich) were deposited at 0.05-0.3\,\AA/s. Cr was used as top electrode material and $\approx$\,40\,nm were deposited at 0.3-1\,\AA/s. The sample was exposed to air for $\sim$\,1\,h between the fabrication steps.
The electrode geometry is optimized to incorporate sub-20\,nm thin Cytop dielectrics in place of much thicker Cytop dielectrics (430-700\,nm)\cite{Kalb2007,Kalb2007a}, i.e. there is no overlap of the gate\,(G) and source/drain\,(S/D) electrodes where they are separated only by the dielectric (metal-insulator-metal region). For PTCDI-C$_{\mathrm{13}}$ TFTs (Fig.\,\ref{fig:trans18nmStat}a), the capacitive G-S and G-D overlap has been reduced to small stripes ($\approx$\,20\,$\mu$m wide), by precise alignment of the shadow masks with specially designed deposition equipment. The channel length $L$ is 250\,$\mu$m and the channel width $W$ is 450\,$\mu$m. For inverters, $L$=\,50\,$\mu$m and $W$=\,400\,$\mu$m (G-S/D overlap is $\approx$\,50\,$\mu$m).
OTFT characteristics were measured with a HP~4155A semiconductor parameter analyzer in a He atmosphere (O$_{\mathrm{2}}$, H$_{\mathrm{2}}$O$<$0.6\,ppm). The integration time was 20\,ms.  The capacitance and electrical insulating properties of Cytop films were measured for metal-insulator-metal (MIM) structures consisting of a Cytop layer sandwiched between Al bottom and Cr top electrodes. Capacitance and leakage current measurements were done in air, with an Agilent 4192A impedance analyzer and an Agilent 4396B high resistance meter.

% include figure 2 -  use [scale=1] for twocolum print
% figure 2

Typical transistor characteristics from OTFTs with an ultra-thin Cytop fluoropolymer dielectric and PTCDI-C$_{\mathrm{13}}$ as semiconductor, are shown in Fig.\,\ref{fig:trans18nmStat}b for three devices (1-3). Particularly noteworthy are several points: low operating voltage with onset at $\approx$\,0\,V, steep subthreshold swing (0.1-0.2\,V/decade), essentially ideal transistor characteristics and extremely small hysteresis, i.e. forward and reverse scans are shown in all panels of Fig.\,\ref{fig:trans18nmStat}, but are indistinguishable. The threshold voltage ($V_{\mathrm{T}}$) was as low as 0.2\,V (Fig.\,\ref{fig:trans18nmStat}c). The steep subthreshold swing, the vanishingly small hysteresis and the low threshold voltage reveal the low interface trap density. The two-terminal mobility determined from transfer characteristics was 0.14\,-\,0.16\,cm$^{\mathrm{2}}$/Vs both for the linear ($V_{\mathrm{DS}}$=\,0.2\,V) and the saturation regime ($V_{\mathrm{DS}}$=\,3.2\,V). Output characteristics, i.e $I_{\mathrm{D}}$ at constant $V_{\mathrm{GS}}$, show linear current increase at low source-drain bias, i.e. ohmic contact resistances, and good saturation at higher bias (Fig.\,\ref{fig:trans18nmStat}d).

The reduction of the gate dielectric thickness from a few 100\,nm to less than 20\,nm results in a substantial increase in the capacitance per unit area and thus in much lower operating voltage. For the substrates of these OTFTs, the measured capacitance per unit area was as high as  100$\pm$5\,nF/cm$^{\mathrm{2}}$ (area 0.49$\pm$0.02\,mm$^{\mathrm{2}}$). Nevertheless, the gate leakage current at 2\,MV/cm is $<$10\,pA if the TFT is operated in the saturation regime (Fig.\,\ref{fig:trans18nmStat}b). Devices of the same geometry and structure exhibit very similar electrical characteristcs.

% include figure 3 - use [scale=1] for twocolum print
% figure 3
\begin{figure}
\includegraphics[scale=1]{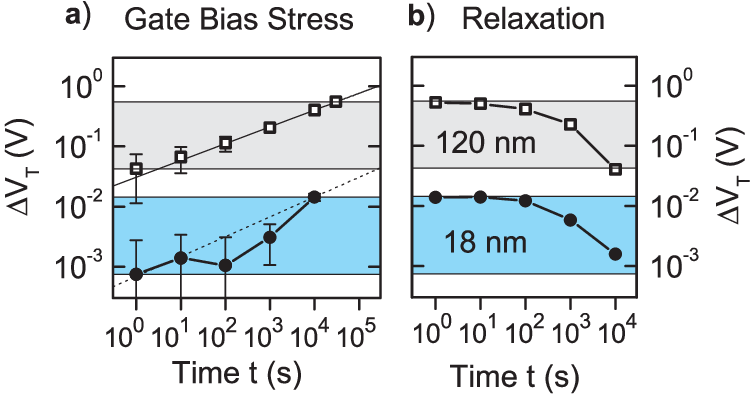}
\caption{\footnotesize\label{fig:stressComp} (Color online) Electrical stability of PTCDI-C$_{\mathrm{13}}$ TFTs: (a)~Threshold voltage shift $\Delta V_{\mathrm{T}}$ due to gate bias stress, for TFTs with  18\,nm thin and 120\,nm thick Cytop gate dielectrics. (b)~Relaxation after gate bias stress (no voltage applied).
}
\end{figure}

Spin-on Cytop intrinsically provides for a high resistance against undesirable deterioration of the as-fabricated transistor characteristics, particularly against gate bias stress\cite{Kalb2007,Kalb2007a,Umeda2008,Walser2009}.
In the following we discuss the electrical stability of PTCDI-C$_{\mathrm{13}}$ TFTs with an 18\,nm thin and for comparison, with an 120\,nm thick Cytop dielectric. The stressing experiment was started with a relaxed device (no voltage applied for several hours). Then, we applied a constant gate bias $V_{\mathrm{G}}$ for uninterrupted time periods of 1,\,9,\,...,\,9\,000 and 20\,000\,s and recorded the transfer characteristics in the linear regime ($V_{\mathrm{DS}}$=\,0.2\,V) after each period. The threshold voltage $V_{\mathrm{T}}$ was determined from fits to straight lines according to
\vspace{-0.4cm} 
\begin{equation}
I_{\mathrm{D}}=WL^{-1}\mu_{\mathrm{lin}}C_{\mathrm{i}}\left(V_{\mathrm{G}}-V_{\mathrm{T}}\right)\,V_{\mathrm{DS}}\quad,
\vspace{-0.1cm} 
\end{equation}

where  $\mu_{\mathrm{lin}}$ denotes the two-terminal mobility. Fig.\,\ref{fig:stressComp}a shows the shift $\Delta V_{\mathrm{T}}$ of the threshold voltage relative to the inital value (0.35\,V for 18\,nm and 2.6\,V for 120\,nm Cytop). The straight line indicates sublinear dependence on the stress time $t$, i.e. $\Delta V_{\mathrm{T}}\propto t^{\beta}$ with $\beta\approx$\,0.3. The gate bias during stress periods was positive (electron accumulation), 2\,V for TFTs with an 18\,nm thin dielectric, and 15\,V for TFTs with an 120\,nm thick dielectric (drain and source voltage were set to 0\,V), i.e. the electric field applied to the dielectric was 1.1\,MV/cm for 18\,nm and 1.25\,MV/cm for 120\,nm Cytop. Because the measurement of the the transfer characteristic involves a gate voltage sweep, its stressing effect is relevant for short stress times, and the estimated effect is indicated by bars in Fig.\,\ref{fig:stressComp}a. We find very small threshold voltage shifts (TVSs) for both, 120\,nm thick and 18\,nm thin Cytop dielectrics. The maximal TVS shown in Fig.\,\ref{fig:stressComp}a, is 0.55\,V for 120\,nm (after 8.3\,h) and as small as 0.014\,V for 18\,nm thin Cytop (after 2.8\,h). The two-terminal mobility was 0.14\,cm$^{\mathrm{2}}$/Vs for TFTs with 120\,nm and 0.13\,cm$^{\mathrm{2}}$/Vs for TFTs with 18\,nm Cytop. Stress measurements were carried out in four-terminal configuration for 120\,nm Cytop. The contact resistance for these top-contact devices was 1-2\,M$\mathrm{\Omega}$ and did not change during gate bias stress. Further studies will clarify the detailed mechanism giving rise to the residual small TVS.

% figure 4 - use [scale=1.1] for twocolumn print
% figure 4
\begin{figure}
\includegraphics[scale=1.1]{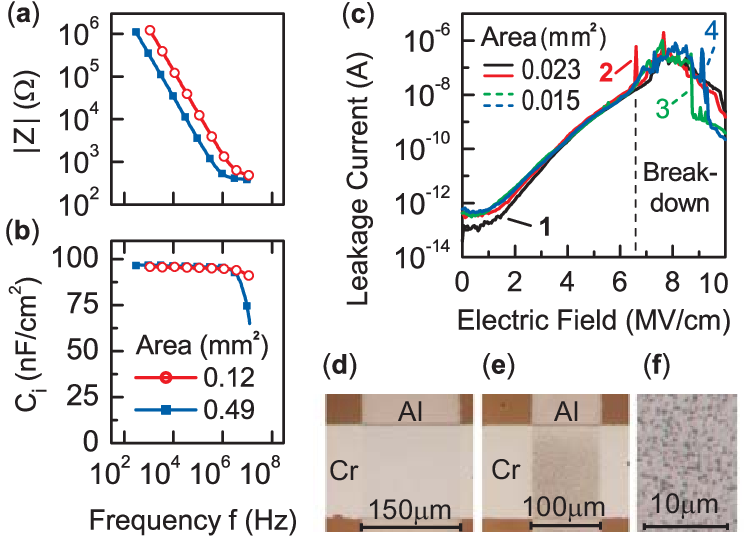}
\caption{\footnotesize \label{fig:cap16nm} (Color online) Performance of 19\,nm thin Cytop dielectrics: (a)~Impedance $Z$ versus frequency $f$ measured on circular MIM structures, used to determine the capacitance $C_{\mathrm{i}}$  per unit area~(b). (c)~Leakage current versus electric field for devices as shown in the optical micrographs~(d,e) with Cr top electrode as deposited~(d), and (e,f)~after electrical breakdown measurement. (f)~Magnification of~(e), showing the uniform distribution of breakdown areas (dark spots).
}
\end{figure}

Given the essential role played by the dielectric layer, we have further characterized the ultra-thin Cytop film by impedance and leakage current measurements on metal-insulator-metal (MIM) structures. The capacitance was determined from impedance measurements according to an equivalent RC circuit (Fig.\,\ref{fig:cap16nm}a). Even for sub-20\,nm thin Cytop layers, the capacitance is essentially frequency independent (Fig.\,\ref{fig:cap16nm}b) and in agreement with the value calculated from the geometry, i.e. $C_{\mathrm{i}}=\varepsilon_{\mathrm{0}}\varepsilon_{\mathrm{i}}d^{-1},$
where $C_{\mathrm{i}}$ is the capacitance per unit area, $\varepsilon_{\mathrm{i}}=\,$2.1 is the permittivity of Cytop and $d$ is the thickness of the insulating layer as measured with a surface step profiler, averaged over several measurement points on the same substrate ($\pm$1\,nm). The high-frequency response ($f>$1\,MHz) is dominated by a serial resistance ($\approx$\,300\,$\mathrm{\Omega}$). Fig.\,\ref{fig:cap16nm}c shows the leakage current for four different devices (1-4) on two substrates, measured with crossed electrodes (Fig.\,\ref{fig:cap16nm}d). The leakage current is well below 10\,pA for electric fields lower than 2\,MV/cm. For these devices, the electrical breakdown occurred above 6\,MV/cm on small spots uniformly distributed over the device area (Fig.\,\ref{fig:cap16nm}e,\,f).

In summary, this study shows that ultra-thin Cytop fluoropolymer gate dielectrics can be successfully incorporated in the fabrication of OTFTs with very low operating voltage. OTFTs with such thin Cytop dielectrics exhibit the same excellent electrical quality and stability as OTFTs with much thicker Cytop dielectrics. These sub-20\,nm thin polymer films form gate dielectrics with high breakdown field and low current leakage, even if spin-coated onto non-flat and rather rough gate electrodes. Thus, the present method to increase the dielectric capacitance and to reduce the OTFT operating voltage significantly expands the options for practical device design and fabrication, and is promising for other material combinations.

We would like to thank K. Pernstich for valuable discussions
and M.P.W. gratefully acknowledges financial support by ETH Zurich through the ESOP scholarship.

\bigskip

\hspace{-0.6cm}
%\center
\begin{minipage}{0.5\textwidth}

\end{minipage}
%\clearpage

\end{document}